\documentclass [prl,amsmath,showpacs,twocolumn,preprintnumbers,superscriptaddress]{revtex4-1}
\usepackage[latin1]{inputenc}% Allow ISO-8859-1 8-bit input
\usepackage{mathtools}
\usepackage{amssymb}
\usepackage{bm}
\usepackage{graphicx}
\usepackage{url}

\RequirePackage{xspace}

\begin{document}

\title{Tachyonic Majorana neutrinos or neutrino spin-to-orbital angular momentum conversion in OPERA}

\author{M. Laveder}
\email{marco.laveder@unipd.it}
\affiliation{Department of Physics and Astronomy, University of Padova, via Marzolo 8, Padova, Italy and INFN - Sezione di Padova}

\author{F. Tamburini}
\email{fabrizio.tamburini@unipd.it}
\affiliation{Department of Physics and Astronomy, University of Padova, via Marzolo 8, Padova, Italy and CIVEN, Venezia}

\begin{abstract}
The new data release of OPERA - CNGS experiment, obtained with a shorter spill of protons \cite{opera2}, confirms the tachyionic behavior expected from the phenomenological model of a Majorana neutrino with a fictitious imaginary mass term acquired during the propagation in the Earth's crust, recently presented by us \cite{tamlav2011}. We performed numerical simulations of neutrino event detections to compare the properties of these Majorana tachyons with the new OPERA results, finding a good agreement. The possibility of spin-to orbital angular momentum conversion that is expected to give a negative squared mass in a medium\cite{tamburini2011storming}, is also briefly discussed.
\end {abstract}


\maketitle

\section{Introduction}

The first set of data (OPERA-1), released on 23 September 2011 by OPERA team, suggested that the muon neutrino could propagate at a speed $v$ larger than of light \cite{opera}. In OPERA experiment, the energetic muonic neutrinos ($\nu_\mu$), mainly produced in the decay, $\pi^{(\pm)} \rightarrow \mu^{(\pm)} +  \nu_\mu \;    (\bar \nu_\mu)$, cross the Earth's crust, a dense and structured medium with variable density $\rho$ in space, $2.7 \leq \rho \leq 3.3$~g/cm$^3$ to reach the Gran Sasso Labs after 735 km travel in $~2.5 \times 10^{-3}$ seconds. This first set of data was made of 16111 events detected in OPERA and the detected events correspond to about $10^{20}$ protons on target collected during the 2009, 2010 and 2011 CNGS runs. 

This superluminal property of the muonic neutrino seems to be confirmed by the new results released on 18 November 2011,
with $(v-c)/c = (2.37 \pm 0.32 \, (stat)^{+0.34}_{-0.24} \, (sys)) \times 10^{-5}$  \cite{opera2}. 
From 21 October to 7 November (here and thereafter OPERA-2), CERN sent a different-shaped neutrino beam to Gran Sasso Labs, made with a much shorter spill of protons to determine with a better precision the time of departure of the neutrinos and verify if the superluminal propagation found in the previous results still occurs.
During a CNGS cycle, a new LHC-type bunched beam made with four bunches, each about $3$~ns (FWHM) long was produced. 
Each bunch was made with $\sim 2.5 \times 10^{11}$ protons and spaced in time by $524$~ns, which means $\sim 1.1 \times 10^{12}$ protons on target for each extraction from the SPS, with the result of decreasing the initial number of protons per cycle of a factor $60$.
The new data show an anticipation with respect to the light time of flight of $62.1 \pm 3.7$~ns, in agreement with the value of $57.8 \pm 7.8 (stat)^{+8.3}_{-5.9} \, (sys)$~ns, obtained with the main new analysis. In the new data analysis, the $5\%$ of spurious data have been discarded and the total statistics used for the analysis was 15223 events that includes 7235 internal, charged and neutral current interactions and 7988 external charged current events, a $70\%$ subsample of the entire OPERA statistics. For better insight see Ref.  \cite{opera2}. 

From simple calculations, one can verify that the relativistic effects due to the different positions of Earth during its orbital motion in the gravitational field of the Sun can be neglected. The effects, expected from the relative velocities from perihelion to aphelion that give a time delay of $2 \times 10^{-9}$ seconds and other relativistic effects that cumulatively give a difference of about $60$~ns, cannot be invoked in OPERA-2 run to explain the anticipation. The Sagnac effect was taken in account too. 

Moreover, some of the experimental conditions changed, as both the spill shape and the experimental conditions, namely the position of the Earth in its orbit and the shorter time of acquisition of the dataset, are slightly different. The only possible source of error that could remain might be hidden 
in the time jitter due to Gran Sasso electronics, which is supposed to be on the order of a few tenth of nanoseconds, that, will be reduced in the next run with a master block of $100$ MHz sampling. In any case also Montecarlo numerical simulations indicated that no instrumental effects on the anticipation $\Delta t$ can be caused by an energy dependent time response of the detector  \cite{opera2}. 

As already discussed by us, this tachyonic behavior of the neutrino seems to emerge only when these quanta propagate inside a material and/or in a gravitational field, where the concentration of sterile neutrinos is expected to increase with respect to deep space \cite{tamlav2011}. A similar interpretation was given when invoking the dependence of neutrino mass on the environmental temperature or energy density \cite{matone}.
In fact, the more stringent limit to their propagation at speed different than light, with $|v-c| / c < 2 \times 10^{-9}$, with the $\bar \nu_e$'s from the supernova (SN) SN1987a \cite{arnett1989supernova} can be reconciled only if we consider a superluminal propagation only inside the SN progenitor, in a path $10^{12} - 10^{13}$~cm long of stellar matter just starting its expansion \cite{hirata1987observation,cullen1999sn}. In the vacuum, the electronic anti-neutrinos, instead, propagated at a speed close to that of light lor the remaining $51.4$ kiloparsecs.
Quantum-gravity (QG) effects  and violations of Lorentz invariance could be held responsible during the interactions of energetic neutrinos  \cite{PhysRevD.77.053014} with space-time fluctuations \cite{ellis2008probes,sakharov2009exploration} but recent results demonstrate that the scales at which quantum gravity phenomenologies emerge are much closer to the Planck scale than those related with OPERA, if the limits estimated for photons apply also to neutrinos \cite{tamburini2011no,PhysRevD.83.121301}. Other aspects, in relation with some scenarios of large-extra-dimension literature have been discussed phenomenologically and proved the consistence of OPERA results with the neutrino data previously obtained at FERMILAB at different energies \cite{camelia}.

In this \textit{letter}, we simulate, with our phenomenological model, based on OPERA-1, MINOS, SN1987a data \cite{tamlav2011}, a set of events and compare them with the new OPERA-2 dataset. With simple logic, one can infer either that  OPERA experiments are affected by a still hidden or underestimated error or that neutrinos actually phenomenologically behave like tachyons when traversing a dense material or a gravitational field. 

\section{Tachyonic Majorana Neutrino}

We now make the hypothesis that both the dataset are free from systematic and interpretation errors, namely that all the delays present in the neutrino production and propagation have been taken in account, together with the synchronization of OPERA and Gran Sasso with GPS (see e.g. Ref. \cite{elburg}).  As claimed in Ref. \cite{tamlav2011}, when crossing a dense medium, neutrinos can behave as real tachyons. Another possibility is a pseudo-tachyonic behavior, similarly to what is observed with photons in a hyperbolic metamaterial \cite{pendry2006controlling,ziolkowski2001superluminal} but it seems to be forbidden by Cohen and Glashow decay model \cite{ceg}  (thereafter, CG). In this case, Standard model (SM) neutrinos are expected not to experience any pseudo-superluminal motion in OPERA experiment. Unavoidably, a disruption of the beam shape due to effects induced by weak-current phenomena should occur, leaving traces of this event in the energy spectrum of the detected neutrinos. In fact, an energetic SM neutrino, traveling faster than light in that medium, is should produce electron/anti-electron pairs radiating away their energy. Evidence of this radiation, emitted from $e^+$ and $e^-$ pairs, was not seen neither during Opera data acquisition \cite{opera} nor with Icarus experiment \cite{icarus}. An important criticism to these anomalous decay processes is that they are forbidden if Lorentz symmetry is instead ``deformed', preserving the relativity of inertial frames by introducing nonlinear terms to energy-momentum relations, as shown in Ref. \cite{camelia2}.

Our model is obtained by applying the tachyonic solution of Majorana infinite--spin component equation to the phenomenological behavior of these neutrinos \cite{Majorana:NC:1932,majorana1937theory}. 
The usual relativistic formulation of mass $m$, momentum $\mathbf{p}$ and energy $W$ of the particle is $W=\sqrt{c^2 p^2 + m^2 c^4}$. 
The solutions to Dirac's equation \cite{Dirac01021928,thaller1992dirac} proposed by Majorana, representing plane waves with positive-defined mass, obey also another class of solutions generated by infinitesimal Lorentz transformations for which $W=\sqrt{c^2 p^2 - m^2 c^4}$ and both the solutions give an energy/angular momentum spectrum that depends on the spin $s$ of the particle, $W_0=m c^2/(s + 1/2)$. Particles with different intrinsic angular momenta then present different masses, determined in the particle's reference frame. 
The tachyonic solution exist for all the positive values of $k$ for which $p \geq kc$ holds. Those states can be considered as belonging to the class of solution with imaginary mass term $\mathrm{i} k$.

This particular phenomenological interpretation of the experimental data is not forbidden by the CG radiation condition.In fact, no radiation is expected from an actual tachyonic behavior, because the standard dispersion relation, that includes the electron-positron pair production, requires that $E_\nu^2-p_\nu^2 > (2m_e)^2$, is not satisfied in the presence of an imaginary mass. The modified dispersion relation $E^2=p^2+m^2+F$ of Ref. \cite{tamlav2011}, where $F$ is an arbitrary function and $p$ the conjugate momentum that depend on space-time coordinates, either because of the medium structure or from the structure of space-time itself, shows a dependence of the energy from $p$ and $F$ and forbids the CG pair production for a neutrino imaginary mass $\sqrt{m^2+F}$, when $F$ is negative. This argumentation finds a correspondence with the introduction of nonlinear or addictional terms in the energy-momentum relationhip of Ref. \cite{camelia2}, with a behavior that is not expected by a SM neutrinos, but can be explained with a particular Majorana neutrino that follows the Majorana mass/spin relationship of 1932 that gives $m^2= - k^2$.
Neutrinos, when traversing layers of matter and/or interacting with sterile neutrinos inside a gravitational field then become tachyonic Majorana neutrinos.
Our phenomenological model of a Majorana neutrino with imaginary mass formulated to explain OPERA anomaly \cite{tamlav2011} shows a good agreement also with the new results. The relative time anticipation $\Delta T / T_0$ and the imaginary mass terms, with uncertainties, obtained with Huzita relationship $m^2=2 E^2 \Delta T / T_0$, \cite{huzita1987neutrino}.

In the first dataset of OPERA-1, MINOS and SN1987a, energy and momentum follow a linear distribution, $m=p_1 E + p_2$.
The fitting parameters for this set of data, together with their $95 \%$ confidence intervals are $p_1 =   0.006727 \;  (0.005509, 0.007945)$ and $p_2 =   0.006884 \;  (0.005824, 0.007945)$~GeV. The latter is the tachyonic mass term of the neutrino obtaned in the limit $p=kc^2$. The sum of squares due to error is SSE=$0.6483$ and the root mean square error is RMSE=$0.4026$ \cite{tamlav2011}. 

Now we simulate with our model a string of $1000$ neutrino events and compare the results with the new release of OPERA data \cite{opera2}.
By applying Montecarlo simulations, we calculate the time anticipation $\Delta T$ of the neutrino signal w.r.t. to the light together with the Majorana imaginary mass, following with the Huzita relationship. The parameter $T_0 = 2.45$~ms is the light time of flight from CERN to Gran Sasso, $m$ is the Majorana imaginary mass term that is supposed to obey the linear distribution $m = p_1 E + p_2$ as a function of the neutrino energy $E$.

The values of the fitted parameters $p_1$ (mean $0.006727$ and std.dev = $0.000609$) and $p_2$ (mean $0.006884$ and std.dev = $0.000530$) are generated according to a normal distribution. The neutrino energies $E$ were generated according to the muon neutrino fluxes at Gran Sasso Labs \cite{Sala} and weighted for the neutrino cross section.
 
In the most conservative approach chosen by OPERA team, the experiment cannot claim a clear energy dependence of $\Delta t$ in the domain explored by OPERA within the accuracy of the measurement performed in the first and second run. To better discard the energy dependence of neutrino anticipation with respect to the speed of light in vacuum further and deeper experimental investigation are needed in a wider range and with a better precision. If we consider, instead, the averaged values of the time anticipations as a function of energy, we find that for the averaged energies $13.8$~GeV and $40.7$~GeV, one finds $\Delta t = (54.7 \pm 18.4 (stat)_{-6.9}^{+7.3}  (sys))$~ns and $\Delta t = (68.1 \pm 19.1 (stat)_{-6.9}^{+7.3}  (sys))$.

%
%% Fig. 1
\begin{figure}[!htb]
\centering
\includegraphics[width=8.8cm, keepaspectratio]{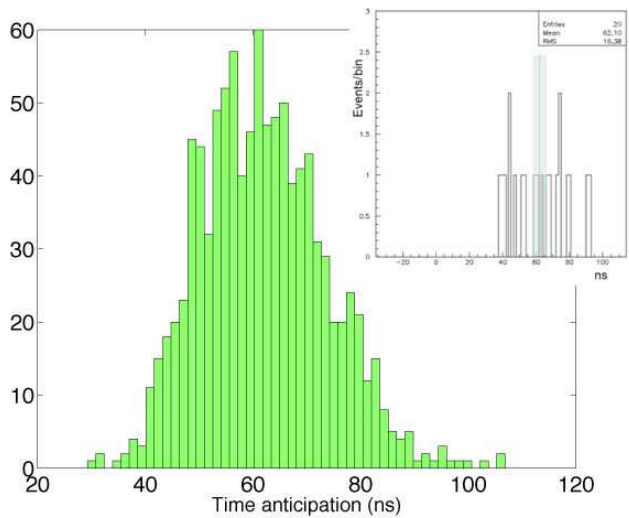}
\caption{Histogram of the time anticipation $\Delta t$  with respect to that of light of the Majorana tachyionic model.
In the inset are reported the new data released by OPERA-2.
The phenomenological model of the tachyonic Majorana neutrino gives an averaged value $\Delta t = 61.95 \pm 0.37$~ns with RMS=$11,73$~ns that well fits with the value $62.1 \pm 3.7$~ns and RMS=$16.4$ reported in the OPERA-2 data. The figure reported in the inset is taken from Ref. \cite{opera2}.}
\label{fig3}
\end{figure}

%
%% Fig. 2 
\begin{figure}[!htb]
\centering
\includegraphics[width=8.8cm, keepaspectratio]{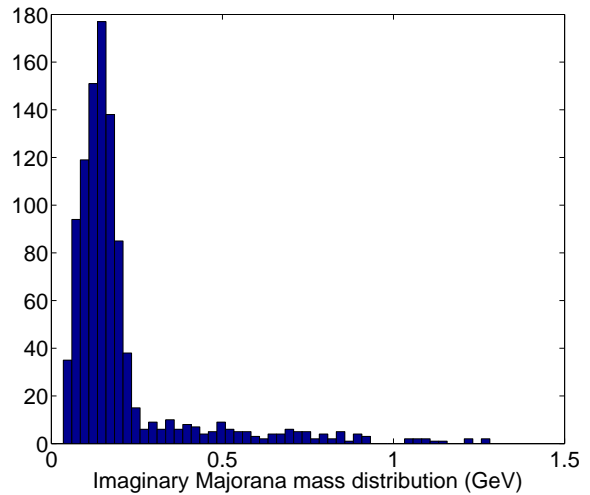}
\caption{Histogram of distribution of imaginary Majorana masses of the neutrino from the Montecarlo simulations of our model, peaked around $0.15$ GeV.}
\label{fig2}
\end{figure}

Most of the simulated mesurements have time anticipations $\Delta T$ in the interval $[40 - 80]$~ns, while the experimental result distribute in an interval $[40 - 90]$~ns, with an averaged value of $61.95 \pm 0.37$~ns with RMS=$11,73$~ns, showing an excellent agreement with OPERA-2 datasets, that gives $62.1 \pm 3.7$~ns with RMS $16.4$~ns, as shown in Fig 1. 
%Moreover, our results show the tipical behavior of a tachionic particle that was obtained neglecting any form of possible sistematic error present in the electronics \cite{opera2} . 
This agreement between OPERA-2 data and our simulations suggests that the possible source of error due to the electronic jitter ($\sim \pm 25$~ns) may have less influence on the experimental data that supposed in Ref. \cite{opera2}. The values of the correspondent imaginary Majorana masses of our model are reported in Fig. 2. Only future measurements will get rid of this residual unknown indetermination.

Another possible cause of superluminality could happen when neutrinos particles interact with a structured-matter medium or when traverse a gravitational field of a rotating black hole \cite{tamburini2011twisting}. In these cases, spin-to-orbital angular momentum (OAM) conversion occurs. A particular medium can exhibit peculiar spatial structures that breaks the space-time symmetry of the space-time manifold structure given by the Lorentz group, in which space is homogeneous and isotropic and time homogeneous \cite{tamburini2011storming,mendonca2008neutrino,tamburini2010photon}.
In this case, the mass/angular momentum relationship is $M_\nu =  m/(\Sigma(\ell,q)+ 1/2)$, where $\Sigma(\ell,q)$ is a general function of the spin $s=1/2$, OAM $\ell$ of the neutrino and with an additional dependence from the characteristic spatial scale of the perturbation $q$. The OAM-induced mass would act as a negative-squared mass term due to the inhomogeneities of the medium and thus giving $m^2= - k^2 ( \Sigma(\ell,q) + 1/2)^2$.

\section{Conclusions}
The recent data released by OPERA experiment seem to confirm the superluminal propagation of muonic neutrinos in the Earth's crust with overall significanve of $6.2 \sigma$ \cite{opera2}.
 
With the new data we validate the phenomenological model of Ref. \cite{tamlav2011}, based on Majorana theory to explain OPERA anomaly. These neutrinos seem to behave like Majorana particles with an imaginary mass induced inside Earth's crust.
As already said, a possible explanation of this behavior can be due to a sterile neutrino mixing confined inside a region where a gravitational field is present or that the presence of matter/gravitational field introduces a preferred reference frame violating CPT symmetry \cite{koch} and/or Lorentz invariance. 
Another cause could be the coupling of neutrinos with structured matter that can give rise to parametric resonances \cite{akhmedov2000parametric}, spin-to-orbital angular momentum conversion, MSW mixing, sterile neutrino states, or due to temperature effects.
%Small changes of the non-inertial forces, due to the periodical motion of Moon-Earth and Sun-Earth, reflect in a change of the beam trajectory in Earth's crust. 
%This should provide a variation of the temperature and local gravitational field distribution along neutrino's path and therefore of their flight time.
The lacking of detection and energy loss from CG effect can be the indirect evidence of a Majorana 
tachyonic neutrino state violating CPT invariance or a spin-to-orbital angular momentum conversion of the neutrino beam, that acts as a negative-squared mass term.

CPT invariance, intimately related with Lorentz invariance violation, is not preserved by Majorana theory \cite{Casalbuoni:2006fa}. 
Some hints of CPT violations have been observed with muonic neutrinos \cite{PhysRevLett.107.021801} and with electronic neutrinos  \cite{PhysRevD.82.113009}. A new generation of experiment are currently being proposed by several groups to study possible CPT violating effects \cite{rubbia,nessie}. For a better insight see the Neutrino Unbound webpage \cite{nu}.
The immediate consequence to the choice of this set of solutions, derived from the set of infinitesimal Lorentz transformations, is that the spectrum of these particles exhibits a relationship between the intrinsic spin angular momentum and the Majorana-mass term, $m$, related to the particle's rest mass or to the acquired virtual mass, $M =m/(s+1/2)$, namely, the particle positive-defined mass value decreases when the spin angular momentum increases. 

%
%These neutrino properties, if confirmed, would not only give crucial information to the astrophysics of supernova explosions, compact objects, stellar interiors and cosmology \cite{1996ApJS}, but revolutionize the standard model of particles \cite{yao2006review} and its extensions to Lorentz-violating phenomenologies \cite{colladay1998lorentz}.

%\acknowledgments
%The authors acknowledge Carlo Giunti, Antonio Masiero, Marco Matone, Massimo Della Valle and Cesare Chiosi  for the helpful discussions.

%


\begin{thebibliography}{}

\bibitem{opera2}%
the OPERA~collaboration, Arxiv preprint  arXiv:1109.4897 - replaced preprint, 2011, submitted to JHEP.

\bibitem{tamlav2011}
F. Tamburini and M. Laveder, Arxiv preprint 1109.5445, 2011

\bibitem{tamburini2011storming}
F. Tamburini and B. Thid{\'e}, EPL \textit{in press}, Arxiv preprint 1105.0700, 2011

\bibitem{opera}%
the OPERA~collaboration, Arxiv preprint 1109.4897v1, 2011

\bibitem{matone}
M. Matone, Arxiv preprint 1111.0270, 2011

\bibitem{arnett1989supernova}
W. Arnett, J. Bahcall, R. Kirshner and S. Woosley, Ann. rev. of Astron. and Astroph., 
\textbf{27}, 629, 1989

\bibitem{hirata1987observation}
K. Hirata et~al., Phys. Rev. Lett. \textbf{58}, 1490, 1987

\bibitem{cullen1999sn}%
S. Cullen and M. Perelstein, Phys. Rev. Lett., \textbf{83}, 268, 1999

\bibitem{PhysRevD.77.053014}
N.~E. Mavromatos et~al.,
Phys. Rev. D, 77, 053014, 2008

\bibitem{ellis2008probes}
J. Ellis et~al., Phys. Rev. D 78 033013, 2008

\bibitem{sakharov2009exploration}
A. Sakharov et~al.,   in Journal of Physics: Conference Series,
171, 012039, 2009

\bibitem{tamburini2011no}
F. Tamburini, C. Cuofano, M. Della Valle, and R. Gilmozzi, A\&A  \textbf{533} A71, 2011

\bibitem{PhysRevD.83.121301}
P. Laurent et~al., Phys. Rev. D, \textbf{83}, 121301, 2011

\bibitem{camelia}
G. Amelino-Camelia et al., Arxiv preprint 1109.5172, 2011

\bibitem{elburg}
R. A. J. van Elburg, Arxiv preprint 1110.2685, 2011

\bibitem{pendry2006controlling}
J. Pendry, D. Schurig  and D. Smith, Science, \textbf{312} 1780, 2006

\bibitem{ziolkowski2001superluminal}%
R. Ziolkowski, Phys. Rev. E,  \textbf{63}, 046604, 2001

\bibitem{ceg}
A.G. Cohen, S.L. Glashow, Arxiv preprint 1109.6562v1

\bibitem{icarus}
the ICARUS~collaboration, Arxiv preprint 1110.3763v1

\bibitem{camelia2}
G. Amelino-Camelia, L. Freidel, J. Kowalski-Glikman, L. Smolin,  Arxiv preprint 1110.0521, 2011

\bibitem{Majorana:NC:1932}
E. Majorana, Nuov.\ Cim.,  \textbf{9}, 335, 1932

\bibitem{majorana1937theory}%
E. Majorana, Nuov.\ Cim.,  \textbf{14}, 171, 1937

\bibitem{Dirac01021928}
P.A.M. Dirac, Proc. R. Soc. Lon. A,  \textbf{117}, 610, 1928

\bibitem{thaller1992dirac}
B. Thaller,  \emph{The Dirac Equation}, Springer, 1992

\bibitem{huzita1987neutrino}
H. Huzita, Modern Physics Letters A, \textbf{2}, 905, 1987
 
\bibitem{Sala}
\url{http://www.mi.infn.it/~psala/Icarus/nugsweb2005/nugs2005numu.flu}
 
\bibitem{tamburini2011twisting}
F. Tamburini, B. Thid{\'e}, G. Molina-Terriza and G. Anzolin, Nature Physics, 
\textbf{7}, 195, 2011

\bibitem{mendonca2008neutrino}
J. Mendon{\c{c}a} and B. Thid{\'e}, EPL, \textbf{84}, 41001, 2008

\bibitem{tamburini2010photon}
F. Tamburini, A. Sponselli, B. Thid{\'e} and J. Mendon{\c{c}a},  EPL, \textbf{90}, 45001, 2010

\bibitem{koch}
B. Koch, Arxiv preprint 1109.5721, 2011

\bibitem{akhmedov2000parametric}
E.  Akhmedov, Pramana, \textbf{54}, 47, 2000.

\bibitem{Casalbuoni:2006fa}
R. Casalbuoni,  PoS {\bf EMC2006} 004, Arxiv preprint hep-th/0610252, 2006.

\bibitem{PhysRevLett.107.021801}%
P.  Adamson et~al., Phys. Rev. Lett., \textbf{107}, 021801, 2011.

\bibitem{PhysRevD.82.113009}%
C. Giunti and M. Laveder, Phys. Rev. D,  \textbf{82}, 113009, 2010.

\bibitem{rubbia}
C. Rubbia et al., CERN-SPSC-2011-012; SPSC-M-773, 2011.

\bibitem{nessie}
P. Bernardini et al., Arxiv preprint arXiv:1111.2242, 2011.

\bibitem{nu}
\url{http://www.nu.to.infn.it/Neutrino_SBL/#20}

%\bibitem{PhysRevD.76.072005}
%P. Adamson et~al., Phys. Rev. D,  \textbf{76}, 072005, 2007
%
%\bibitem{Stal&al:PRL:2007}
%O. St{\aa}l et~al., Phys. Rev. Lett., \textbf{98}, 071103, 2007
%
%\bibitem{akhmedov2001floquet}
%E. Akhmedov, Physics of Atomic Nuclei, \textbf{64}, 787, 2001
%
%\bibitem{petschek1990supernovae}%
%  \emph{\bibinfo {title} {Supernovae}},\ edited by A. Petschek
%  New York, NY (USA); Springer-Verlag New York Inc., 1990
%
%\bibitem{woosley1986physics}
%S. Woosley and T. Weaver, Ann. rev. of astron. and astroph., \textbf{24}, 205, 1986
%
%\bibitem{smolyaninov2011metamaterial}
% I. Smolyaninov, Journal of Optics,  \textbf{13}, 024004, 2011
% 
%\bibitem{1996ApJS}
%N. Itoh, H. Hayashi, A. Nishikawa and Y. Kohyama, ApJS,  \textbf{102},  411, 1996
%
%\bibitem{yao2006review}
%W. Yao et~al., Journal of Physics G, \textbf{33}, 1, 2006
%
%\bibitem{colladay1998lorentz}%
%D. Colladay and V. Kosteleck{\`y}, Phys. Rev. D, \textbf{58}, 116002, 1998


\end{thebibliography}
\end{document}